\documentclass[twocolumn]{iopart}
\usepackage{epsfig,iopams}
\newcommand{\Jm}{J_{\mathrm{max}}}
\newcommand{\lm}{\lambda_{1}}

\begin{document}

\title{Lyapunov instabilities in lattices of interacting classical spins at infinite temperature}
\author{A. S. de Wijn}
\address{Department of Physics, Stockholm University, 106 91 Stockholm, Sweden}
\ead{dewijn@fysik.su.se}
\author{B. Hess}
\address{Institute for Theoretical Physics, University of Heidelberg, Philosophenweg 19, 69120 Heidelberg, Germany}
\author{B. V. Fine}
\address{Institute for Theoretical Physics, University of Heidelberg, Philosophenweg 19, 69120 Heidelberg, Germany}
\ead{B.Fine@thphys.uni-heidelberg.de}

\begin{abstract}
We numerically investigate Lyapunov instabilities for one-, two- and three-dimensional lattices of interacting classical spins at infinite temperature. We obtain the largest Lyapunov exponents for a very large variety of nearest-neighbor spin-spin interactions and complete Lyapunov spectra in a few selected cases. We investigate the dependence of the largest Lyapunov exponents and whole Lyapunov spectra on the lattice size and find that both quickly become size-independent. Finally, we analyze the dependence of the largest Lyapunov exponents on the anisotropy of spin-spin interaction with the particular focus on the difference between bipartite and nonbipartite lattices.

\end{abstract}

\maketitle 

\section{Introduction}

Investigations of the Lyapunov instabilities in many-particle systems are often motivated by the role of chaos in the foundations of statistical physics~\cite{Gibbs-02,Krylov-79,Gaspard-98}, which is still not fully understood.
In general, interacting many-particle classical systems are expected to be chaotic.
The largest Lyapunov exponents and properties of the entire Lyapunov spectra have been calculated numerically for classical many-body systems, such as gases of hard-core particles~\cite{posch1,posch2}, fluids with soft interactions~\cite{soft}, and lattice two-dimensional rotators~\cite{soft,ruffo,Vallejos-02,Anteneodo-03,Vallejos-04}, and analytically in a few cases~\cite{prlramses,lagedichtheid,astridhenk,mareschal,cylinders,roc_astrid_henk,barnett,Vallejos-02,Anteneodo-03,Vallejos-04}.
In the present paper, we focus on lattices of interacting classical spins.

Classical spins often appear in the theoretical studies as the large-spin limit of quantum spins.
Moreover, even when one deals with lattices of spins 1/2, parallels between classical and quantum spin dynamics still remain.
These parallels have recently received much attention, in particular, in the context of the asymptotic exponential-oscillatory behavior of nuclear spin decays in solids~\cite{Fine-04,Fine-03,Fine-05,Morgan-08,Sorte-11,Sorte-12,Meier-12}.
These decays were identified in Refs.~\cite{Fine-04,Fine-05} with chaotic eigenmodes in both classical and quantum many-spin systems.

Although it appears very likely {\it a priori} that lattices of interacting classical spins exhibit chaotic dynamics, no systematic investigation of the chaotic properties of these lattice was undertaken until our previous work~\cite{prlchaosspins}, which presented a survey of the largest Lyapunov exponents for a very large variety of spin lattices and Hamiltonian anisotropies.
The principal finding of Ref.~\cite{prlchaosspins} was that all Hamiltonians considered, with the exception of the Ising case, led to chaotic dynamics as evidenced by the nonzero value of the largest Lyapunov exponent.
We also obtained both analytically and numerically the power-law scaling of the largest Lyapunov exponent in the vicinity of the integrable Ising limit.

In the present paper, we complement the findings of Ref.~\cite{prlchaosspins} in several respects.
Namely, we compute complete Lyapunov spectra for a few selected spin lattices and show their dependence on the lattice size, and also present a more extensive investigation of the lattice size dependence of the largest Lyapunov exponent.
Finally, we discuss the dependence of the largest Lyapunov exponent on the Hamiltonian anisotropy with particular emphasis on the difference between bipartite and nonbipartite lattices.

\section{General formulation}

\subsection{Spin model}

We consider periodically closed spin lattices with the nearest-neighbor (NN)  interaction Hamiltonian of the following kind: 
\begin{equation}
H=\sum_{i<j}^{\hbox{\scriptsize NN}} (J_x S_{ix} S_{jx}+J_y S_{iy} S_{jy}+J_z S_{iz} S_{jz})~,
\label{H}
\end{equation}
where $(S_{ix}, S_{iy}, S_{iz}) \equiv {\mathbf S}_i$ are the three projections of the  classical spin vector of unit length on the $i$th lattice site (i.e.\ ${\mathbf S}_i^2 = 1$), and $J_x$, $J_y$, $J_z$ are the coupling constants, which we also normalize by condition $J_x^2 +J_y^2 + J_z^2 = 1$. Below, we often mention Ising, Heisenberg and ``anti-Heisenberg'' limits of the Hamiltonian (\ref{H}). The Ising Hamiltonian corresponds to $J_x = J_y = 0$, $J_z = 1$, Heisenberg Hamiltonian to  $J_x = J_y = J_z = 1/\sqrt{3}$, and, finally, anti-Heisenberg Hamiltonian to $J_x = J_y = - J_z= 1/\sqrt{3}$. The total number of spins in the lattice is denoted as $N$. The phase space of such a lattice has dimensionality $2 N$. 

\begin{figure}
\epsfig{figure=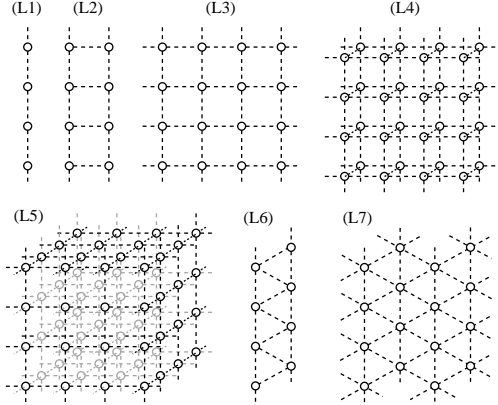,width=6.6cm}
\caption{Lattices investigated in this work.  Bipartite: (L1) chain, (L2) rectangular ladder, (L3) square lattice, (L4) bilayer of square lattices, (L5) cubic lattice.
Non-bipartite: (L6) triangular ladder, (L7) triangular lattice. 
\label{fig:lattices} 
}
\end{figure}

We consider seven lattices shown in Fig.~\ref{fig:lattices} and labeled as (L1-L7). Lattices (L1-L5) are bipartite, which means that they can be divided into two sublattices such that all the interacting neighbors for a spin on one sublattice belong to the other sublattice. Lattices (L6,L7) are nonbipartite. 

The equations of motion associated with the Hamiltonian~(\ref{H}) can be obtained in the Poisson-bracket formalism~\cite{steiner,yang,bensthesis}: $d S_{i,\mu} /dt = \{ {\cal H}, S_{i,\mu} \}$, where the index $\mu$ admits values $1$, $2$ or $3$ representing the projections $x$, $y$, or $z$, respectively. The primary Poisson brackets are: $ \{ S_{i,\mu}, S_{j,\nu} \} = \delta_{ij} \sum_{\kappa} \epsilon_{\mu \nu \kappa} S_{i,\kappa}$, where $\delta_{ij}$ is the Kronecker delta and $\epsilon_{\mu \nu \kappa}$ the Levi-Civita symbol.
The evaluation of $\{ {\cal H}, S_{i,\mu} \}$ involves the following Poisson brackets:
$\{ S_{j,\nu} S_{k,\sigma}, S_{i,\mu} \} = \{ S_{j,\nu}, S_{i,\mu} \} S_{k,\sigma} + S_{j,\nu} \{S_{k,\sigma}, S_{i,\mu} \}$.
The resulting equations of motion are
\begin{equation}
\dot{\mathbf S}_i = {\mathbf S}_i \times {\mathbf h}_i~,
\label{eq:eom}
\end{equation}
where ${\mathbf h}_i$ is the local field given by the expression
\begin{equation}
{\mathbf h}_i = \sum_{j(i)} (J_x S_{jx} {\mathbf e}_x + J_y S_{jy} {\mathbf e}_y + J_z S_{jz} {\mathbf e}_z)~.
\label{h}
\end{equation}
Here ${\mathbf e}_x$, ${\mathbf e}_y$ and ${\mathbf e}_z$ are the unit vectors along the respective directions, and $j(i)$ implies the summation of the nearest neighbors of the $i$-th lattice site.

In this work, we restrict ourselves to the Lyapunov instabilities on the zero energy shell, which corresponds to infinite temperature in the microcanonical sense.

\subsection{Analytical considerations}
\label{Expectations}

Since the dynamics of this system are fully time-reversible, positive and negative Lyapunov exponents are expected to form conjugate pairs with equal absolute values.
In the general case of different $J_x$, $J_y$, and $J_z$, there should be only two zero Lyapunov exponents, corresponding to the energy and the time shift directions.
In the case $J_x=J_y \neq J_z$, of which the anti-Heisenberg case is an example, the $z$-component of the total spin polarization becomes an integral of motion, and hence one more pair of Lyapunov exponents should also become equal to zero.
In the Heisenberg case, $J_x=J_y = J_z$, all three components of the total spin polarization become integrals of motion.
However, these three integrals of motion are not dynamically independent, because one of them can be obtained as the Poisson bracket of the two others. The two independent integrals of motion thus imply two additional pairs of zero Lyapunov exponents. In the Ising case, the $z$-component of each spin is an integral of motion. Hence, the system is integrable and all Lyapunov exponents are equal to zero. 

Now we discuss the connections between the Lyapunov spectra of different anisotropic Hamiltonians. We first note that infinite-temperature Lyapunov spectra are expected to be identical for the Hamiltonians with the coupling constants $(J_x, J_y, J_z)$ and $(-J_x, -J_y, -J_z)$, because the zero-energy shells in the two cases are identical, while the change of the sign of the coupling constants amounts to the operation of time-reversal
and flips the sign of the energy.  
We further note that, for bipartite lattices, the
infinite temperature
Lyapunov spectra for the coupling constants $(J_x, J_y, J_z)$ and $(-J_x, -J_y, J_z)$ should also be identical. In order to see this, one should make the transformation $S_{ix} \to - S_{ix}$ and $S_{iy} \to -S_{iy}$ for one of the two sublattices forming the bipartite lattice and then examine the resulting equations of motion. 

The two observations made above also imply that, for a given bipartite lattice, the Lyapunov spectra for the Heisenberg and the anti-Heisenberg Hamiltonians are identical to each other, which means that, in the latter case, the Lyapunov spectrum has three pairs of zero Lyapunov exponents instead of two expected for the generic case of $J_x = J_y \neq J_z$. The extra pair of the zero exponents originates from the following two (not independent) integrals of motion: $\sum_i (-1)^{\xi} S_{ix}$ or $\sum_i (-1)^{\xi} S_{iy}$, where $\xi$ is the index taking value 0 for one sublattice and 1 for the other. 
The same considerations also imply that any Hamiltonian with $J_x = -J_z$, or equivalent, on a bipartite lattice can be converted to the axially symmetric Hamiltonian with $J_x = J_z$. Therefore, the original Hamiltonain has an extra pair of zero Lyapunov exponents corresponding to the integral of motion $\sum_i (-1)^{\xi} S_{iy}$.

Small spin clusters may have additional nontrivial integrals of motion. 
In particular, any 4-spin periodic chain with the general anisotropic Hamiltonian of form (\ref{H}) is fully integrable. 
This is because the first and third spins in this chain rotate in the same local field $\left[ J_x (S_{2x} + S_{4x}), \ J_y (S_{2y} + S_{4y}), \ J_z (S_{2z} + S_{4z})   \right]$, while the second and the fourth spins rotate in another local field $\left[J_x (S_{1x} + S_{3x}), \ J_y (S_{1y} + S_{3y}), \ J_z (S_{1z} + S_{3z})   \right]$.
Therefore, there are two additional integrals of motion, namely: ${\mathbf S}_1 \cdot {\mathbf S}_3$ and ${\mathbf S}_2 \cdot {\mathbf S}_4$. Finally, one can also check that  $({\mathbf S}_1 + {\mathbf S}_3) \cdot ({\mathbf S}_2 + {\mathbf S}_4)$ is also an integral of motion. As a result, the number of integrals of motion (including energy) becomes 4, while the dimensionality of the phase space is 8, i.e. the problem is fully integrable. In the case of the isotropic Heisenberg Hamiltonian, a much larger variety of small spin clusters with nontrivial integrals of motion were cataloged in Ref.~\cite{Steinigeweg-09}.

\subsection{Numerical simulations}

The equations of motion were integrated using a fourth-order Runge-Kutta algorithm with a time-step $\delta t = 0.005$.
This is sufficiently small so that on the time scale of our simulations, energy is conserved with the 6-digit accuracy.
Simulations were run for 20000 time units, which was sufficient for accurate convergence of the smaller exponents.

The Lyapunov exponents were obtained by using a version of the standard reorthonormalization algorithm~\cite{posch1,Benettin-80,Elsayed-11}.
In order to obtain the first $n$ largest Lyapunov exponents, we numerically propagate a reference trajectory ${{\bgamma}}(t)$ and  $n$ initially orthogonal perturbation vectors $\delta{\bgamma}_i(t)$.
At every time step, we propagate the perturbations using the linear tangent space map that is obtained by numerically taking derivatives,
\begin{equation}
\delta {\bgamma} (t+\delta t) = \frac{\partial {\bgamma}(\delta t)|_{{\bgamma}(0)={\bgamma}_0}}{\partial {\bgamma}_0} \delta{\bgamma}(t)~.
\end{equation}
After each time interval $\Delta t=0.25$, we hierarchically reorthogonalize the perturbation vectors $\delta{\bgamma}_i(t)$ using the Gram-Schmidt procedure, and then renormalize their lengths back to the initial values. The renormalization factors are recorded for the $k$th time interval $\Delta t$ as $\alpha_i(k)$. Finally, we compute the $i$th Lyapunov exponent using the formula $\lambda_i = {1 \over K \Delta t} \sum_{k=1}^{K} \ln \alpha_i(k)$, where $K$ is the number of the renormalization time intervals. 

As a test of the accuracy of our numerical routine, we have checked that the Lyapunov exponents form conjugate pairs and that the number of zero Lyapunov exponents is equal to the number expected from the symmetry of the Hamiltonian as discussed in Section~\ref{Expectations}.  We have also checked the symmetries between Heisenberg and anti-Heisenberg Lyapunov spectra expected based on the arguments presented in Section~\ref{Expectations}.
 
In order to obtain initial conditions at zero total energy, we first choose random orientations for all spins.
This produces a total energy close to zero, but with fluctuations of the order of $\sqrt{N}$.
Then, in order to arrive at zero total energy, we evolve the dissipative equations of motion:
\begin{equation}
\dot{\mathbf S}_i =\pm {\mathbf S}_i \times ({\mathbf S}_i \times {\mathbf h}_i)~.
\end{equation}
Depending on the sign in front of the right-hand side, this increases or decreases the total energy associated with the Hamiltonian~(\ref{H}).
Once the zero value of the total energy is reached, we additionally assure the randomness of the initial conditions on the energy shell by performing $10N$ sequential rotations of random spins by random angles around the directions of their respective local fields ${\mathbf h}_i$ given by Eq.(\ref{h}). These rotations preserve the total energy.

\section{Results and discussion}

\subsection{Lyapunov spectra}

\begin{figure}
\epsfig{figure=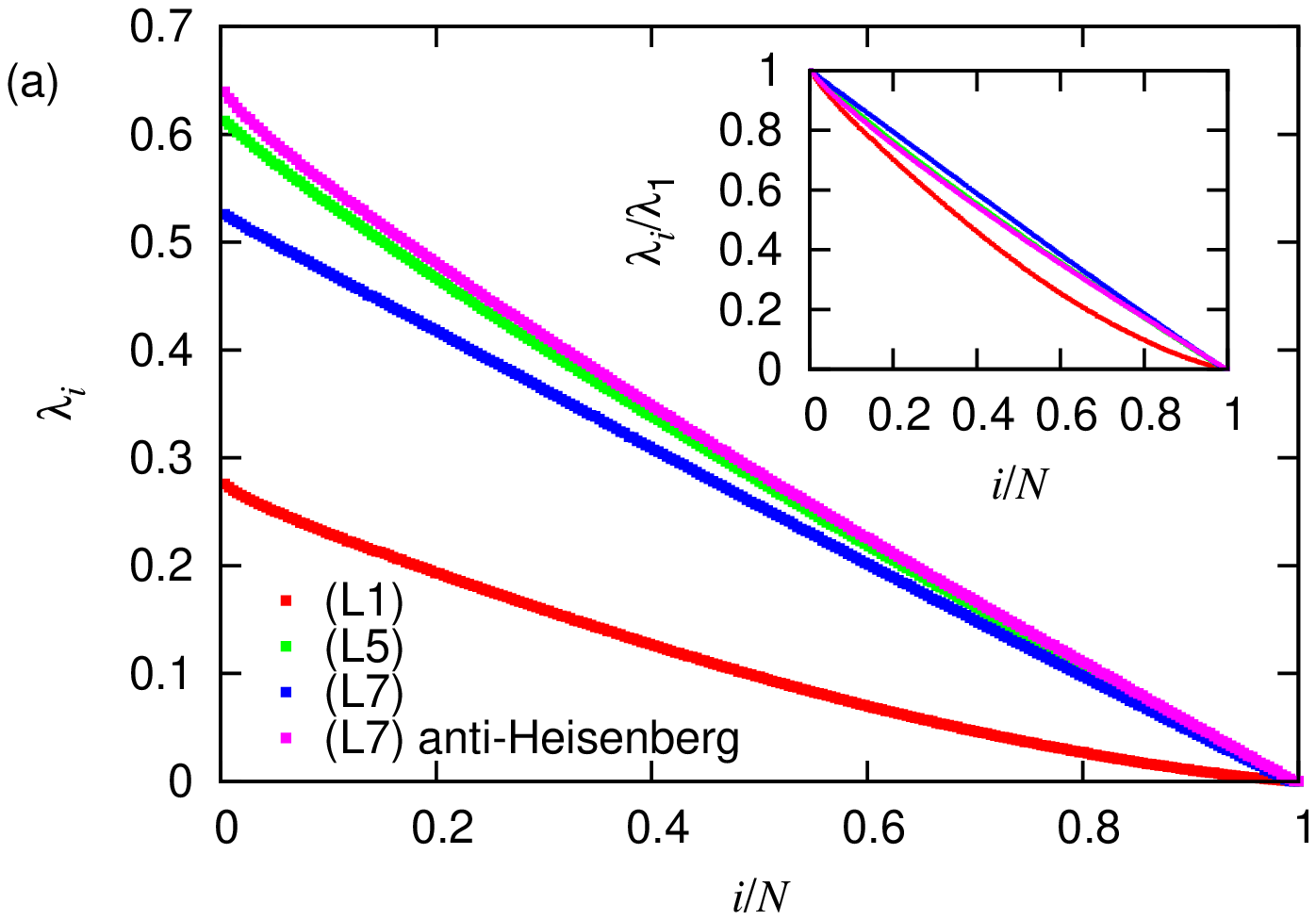,angle=270,width=6.6cm}
\epsfig{figure=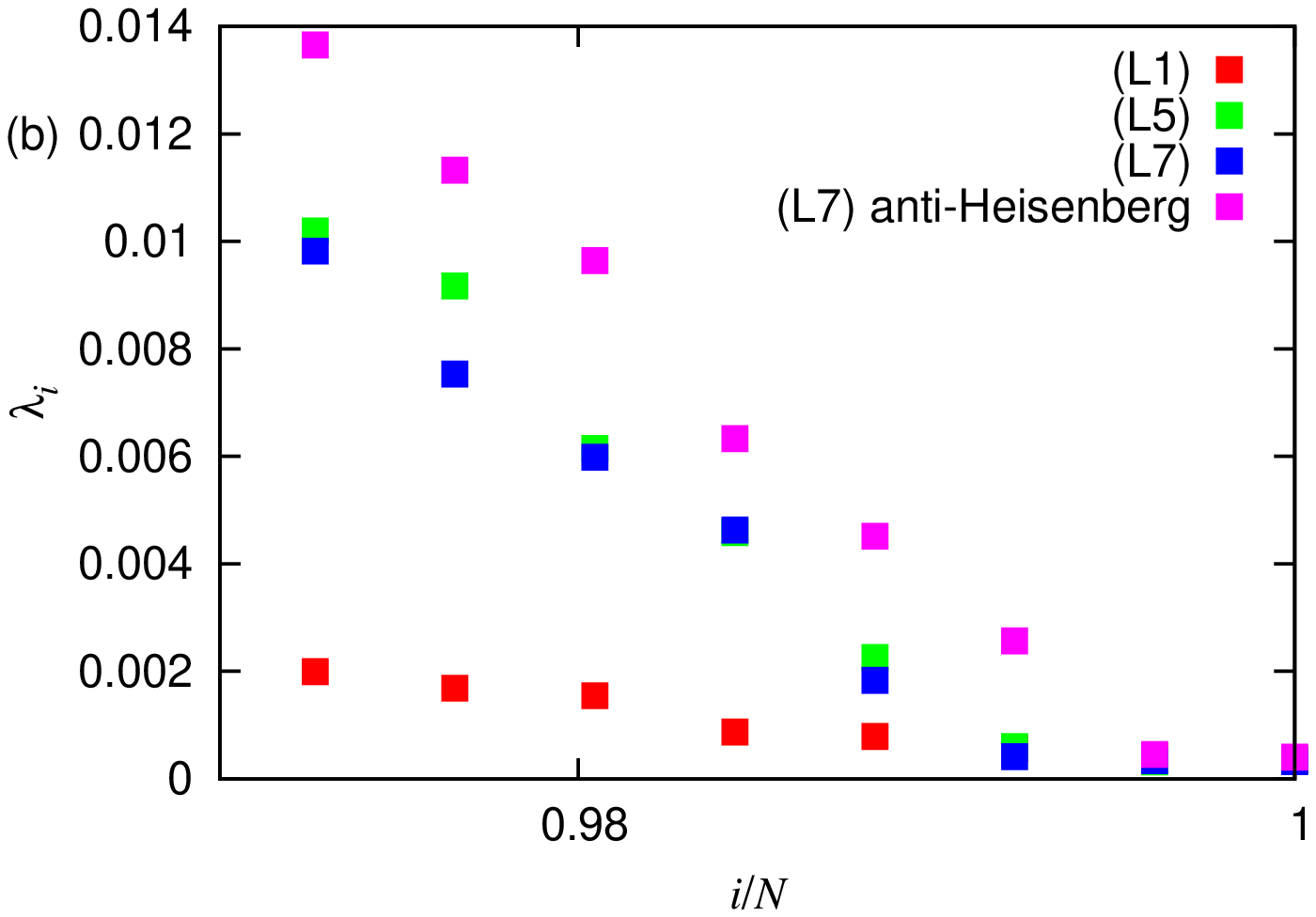,angle=270,width=6.6cm}
\caption{
(Color online) Lyapunov spectra for lattices (L1), (L5), and (L7) with Heisenberg and anti-Heisenberg Hamiltonians. The number of spins in each case is $N=256$. Lyapunov exponents $\lambda_i$ are ordered according their values and plotted as a function of $i$:  (a) Positive halves of the spectra. The inset shows the same plots but with the vertical axis renormalized by the value of the largest Lyapunov exponent $\lambda_1$.  (b) Magnified small-exponent parts of the spectra.
Note: lattices (L1) and (L5) have identical Lyapunov spectra for the Heisenberg and the anti-Heisenberg interactions (see Section~\ref{Expectations}).  In (b), the numerical error is roughly the same as the size of the symbols.
\label{fig:spectraheis}
}
\end{figure}

\begin{figure}
\epsfig{figure=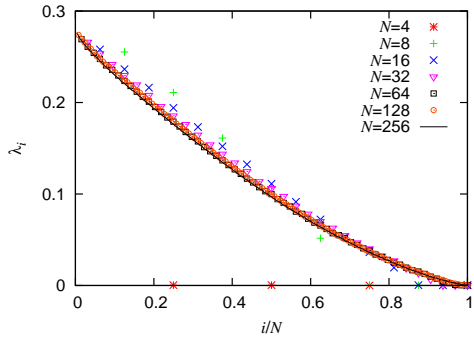,angle=270,width=6.6cm}
\caption{
(Color online) The positive halves of the Lyapunov spectra for spin chains (L1) of different lengths $N$ with the Heisenberg Hamiltonian.
Note: for $N=4$, all Lyapunov exponents are zero, due to the symmetries in the system (see Section~\ref{Expectations}). 
}
\label{fig:sizespectrum}
\end{figure}

We have computed the full Lyapunov spectra for the linear chain (L1), cubic lattice (L5) and triangular lattice (L7) with the Heisenberg Hamiltoninan, and also for the triangular lattice (L7) with the anti-Heisenberg Hamiltonian.  The results are presented in Fig.~\ref{fig:spectraheis}. 
The spectra (i.e.\ the Lyapunov exponents as a function the exponent's index) are typically weakly convex, regardless of the lattice.
In the case of a cubic lattice with Heisenberg interaction, the spectrum is actually very close to linear. As explained in Section~\ref{Expectations}, the Lyapunov spectra for the bipartite lattices (L1) and (L5) with anti-Heisenberg Hamiltonian are identical to those already presented in Fig.~\ref{fig:spectraheis} for the Heisenberg Hamiltonian.
The effect of the lattice size on the Lyapunov spectrum for the linear chain with Heisenberg Hamiltonian is shown in Fig.~\ref{fig:sizespectrum}.
The size dependence of the spectrum appears to be very weak for $N>4$ and becomes virtually unobservable for chains containing more than 32 spins.

Comparing the Lyapunov spectra of classical spins with the Lyapunov spectra of other many-particle systems, we first note, that the spectra presented in Fig.~\ref{fig:spectraheis} do not exhibit an offset from zero for the smallest positive exponents. Such an offset was observed in gases of particles~\cite{posch1} and high-dimensional billiards~\cite{onslorentz,cylinders} but not in systems with sufficiently soft interactions~\cite{soft}. The classical spin lattices obviously belong to the latter group.

We further remark on the existence of the delocalized Lyapunov-Goldstone modes, which were observed in dilute gases~\cite{posch1,taniguchi05,onszelf,mareschal} and in some other extended systems~\cite{HongLiuPRL2008}. If these modes exist in the spin systems, the projections on single spins of the Lyapunov vectors corresponding to the smallest nonzero Lyapunov exponents should exhibit a sinusoidal dependence on the positions of spins. In the present work, we did not investigate the properties of the Lyapunov vectors systematically. However, our several attempts to find the Lyapunov-Goldstone modes for the infinite temperature energy shells did not produce any positive evidence: the projections of the Lyapunov vectors for all exponents were strongly localized on the lattice and not sinusoidal. There is also no indication of a dependence of the smalles non-vanishing exponent on the system length, as is characteristic for the Lyapunov-Goldstone modes.  The Lyapunov-Goldstone modes may still exist in classical spin systems at low temperatures but this is a subject beyond the scope of the present paper.

\subsection{Largest Lyapunov exponents: finite size effects}

Accurate numerical calculations of full Lyapunov spectra are very demanding, because the computational cost grows as $N^2$. For this reason, the lattices investigated in the preceding subsection were relatively small.
In this subsection, we focus only on the largest Lyapunov exponents, $\lm$.
The cost of computing $\lm$ grows only as $N$, which allows us to investigate much larger lattices and a much greater variety of Hamiltonians. An extensive investigation of this kind was already reported by us in Ref.~\cite{prlchaosspins}. In the present and the next subsections we present some results and analysis that were not included in Ref.~\cite{prlchaosspins}.

\begin{figure}
\epsfig{figure=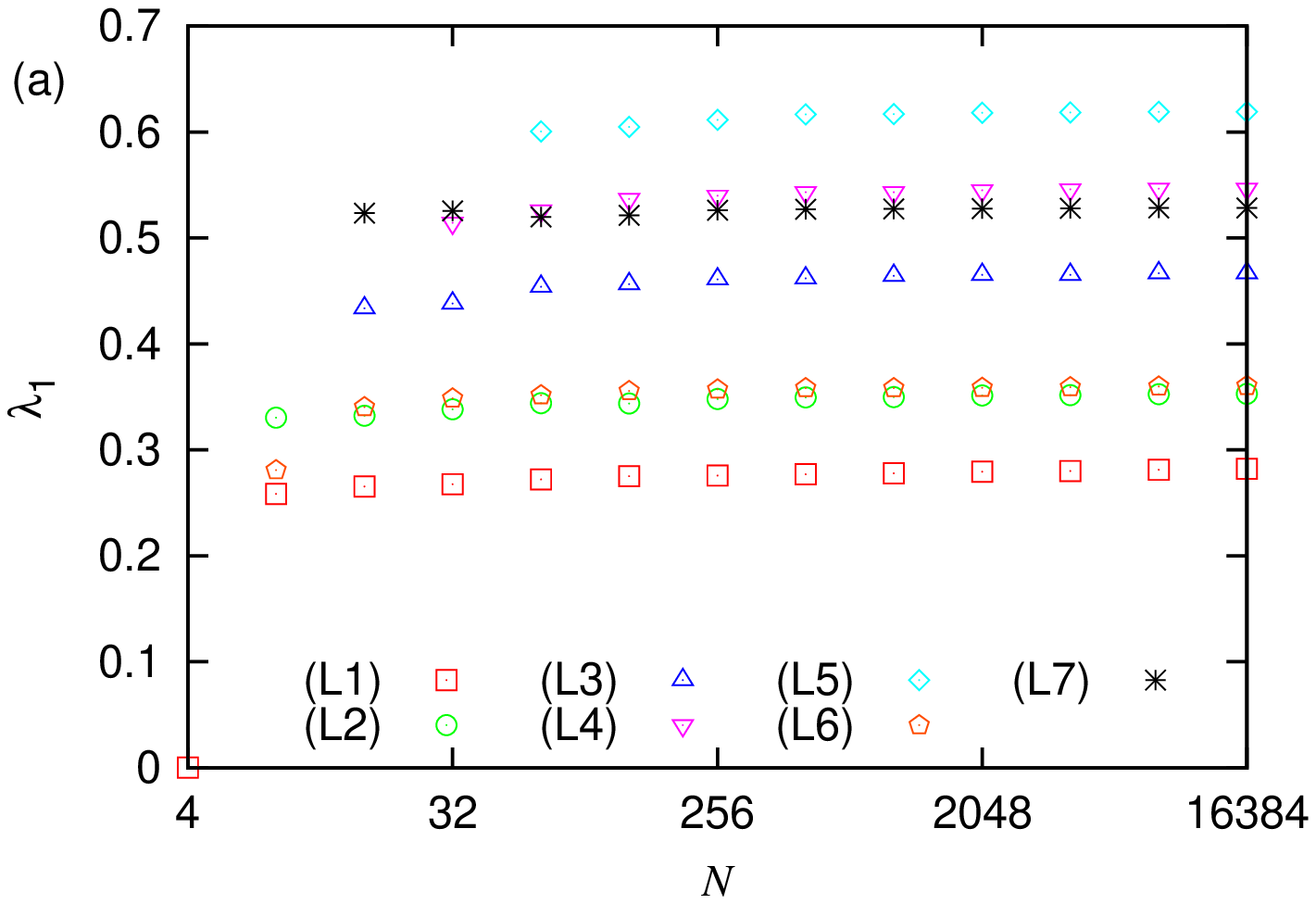,angle=270,width=6.6cm}
\epsfig{figure=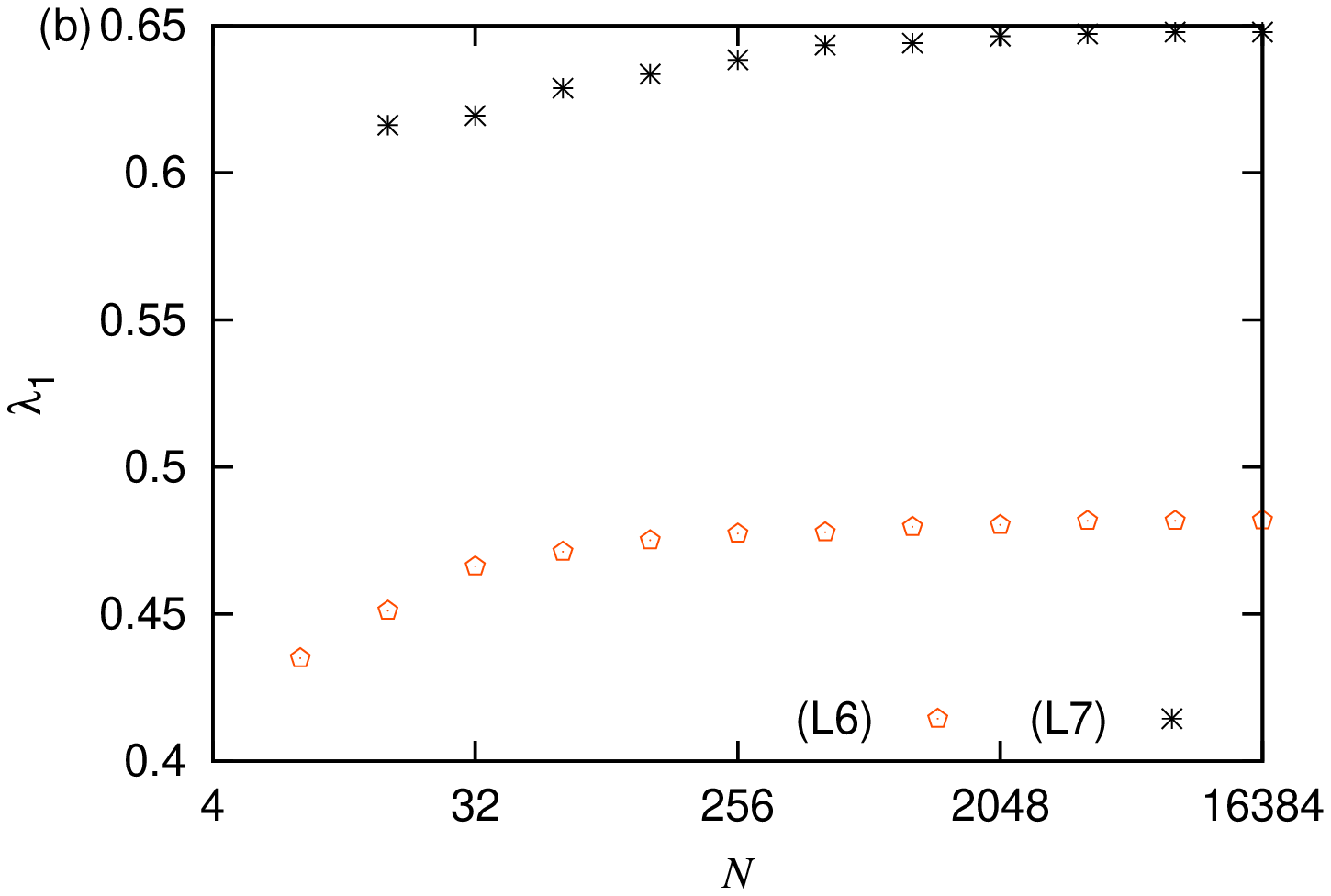,angle=270,width=6.6cm}
\caption{\label{fig:sizedependence}
(Color online) The dependence of the largest Lyapunov exponent on the number of spins $N$ for each of the seven lattices indicated in the plot legent for (a) Heisenberg coupling and (b) anti-Heisenberg coupling.  For bipartite lattices, the largest Lyapunov exponents are the same for Heisenberg and anti-Heisenberg coupling (see Section~\ref{Expectations}).
The largest Lyapunov exponents become constant for sufficiently large systems.
}
\end{figure}

In this subsection, we investigate the dependence of $\lm$ on the lattice size for all seven lattices shown in Fig.~\ref{fig:lattices} with Heisenberg and the anti-Heisenberg Hamiltonians.
The results are shown in Fig.~\ref{fig:sizedependence}. They indicate that, within our numerical accuracy, $\lm$ becomes size-independent for sufficiently large lattices.
In particular, on the basis of these results even slow logarithmic growth of $\lm$ with the lattice size can be excluded.

The saturation of $\lm$ with the lattice size can be explained on the basis of the following consideration. The exponential growth of the perturbation vector ${\delta\bgamma}_1$ in many-spin phase space with rate $\lm$ implies that the projection ${\delta\bgamma}_1$ on the subspace of each individual spin $\{ S_{ix}, S_{iy}, S_{iz} \}$ should also, on average, grow exponentially with the same rate.
The instantaneous growth rate of the perturbations of the coordinates of a given spin can be obtained from the linearized equations of motion, which, in turn can be obtained from equations~(\ref{eq:eom}) and~(\ref{h}),
\begin{eqnarray}
\delta \dot{\mathbf  S}_i = \delta {\mathbf  S}_i \times {\mathbf h}_i + {\mathbf S}_i \times  \delta {\mathbf  h}_i~,\\
\delta {\mathbf  h}_i = \sum_{j(i)} J_x \delta S_{jx} {\mathbf e}_x + J_y \delta S_{jy} {\mathbf e}_y + J_z \delta S_{jz} {\mathbf e}_z~.
\end{eqnarray}
Here, ${\mathbf \delta S_i}$ indicates a perturbation in the $i$-th spin, while $\delta S_{ix}$ indicates a perturbation in its $x$th component.
From these equations, one can see that the instantaneous growth rate is limited from above by the value on the order of the maximum value of the local field $\max(|{\mathbf h}_i|) =  n_0 \ \max(|J_x|, |J_y|, |J_z|)$, where $n_0$ is the number of nearest neighbors for each lattice site.
Since this constraint does not depend on the lattice size, the growth of the perturbation belonging to the largest Lyapunov exponent must saturate as the lattice size increases.
In principle, it might also be possible for the largest Lyapunov exponent to oscillate with the lattice size, but in a system with exponential decay of spatial correlation, this is extremely unlikely.

\subsection{Largest Lyapunov exponents: dependence on the Hamiltonian anisotropy}

\begin{figure}
\epsfig{figure=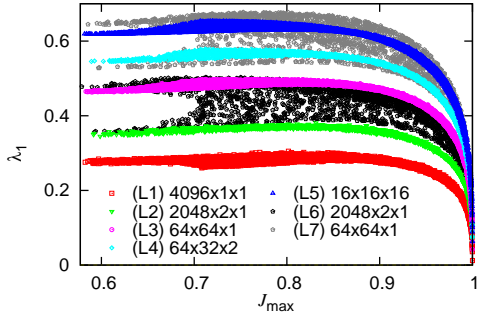,angle=270,width=6.6cm}
\caption{
(Color online) Largest Lyapunov exponents as presented in Ref.~\cite{prlchaosspins}. Each point represents one $\lm$ obtained numerically for a lattice indicated in the plot legend with one randomly chosen set of values $J_x$, $J_y$ and $J_z$ as described in the text.
\label{fig:prl}
}
\end{figure}

In Ref.~\cite{prlchaosspins}, we conducted a systematic survey  of the dependence of $\lm$ on the Hamiltonian anisotropy on the ``interaction sphere'' constrained by the condition $J_x^2 +J_y^2 + J_z^2 = 1$. We have found that the principal parameter controlling this dependence is $\Jm \equiv \max(|J_x|, |J_y|, |J_z|)$. 
In particular, this parameter quantifies the approach to the integrable Ising case corresponding to $\Jm = 1$. The main plot of Ref.~\cite{prlchaosspins} is reproduced in Fig.~\ref{fig:prl}. 

Here we focus on the difference between the anisotropy dependence of $\lm$ for bipartite lattices (L1-L5) and nonbipartite lattices (L6, L7). As can be seen in Fig.~\ref{fig:prl}, the bipartite lattices (L1-L5) show a nearly universal dependence $\lm(\Jm)$, which scales only with the number of interacting neighbors. The small spread of the sampled values of $\lm$ at a given value of $\Jm$ indicates that $\lm$ depends only very little on the ratio of the two coupling constants that have smaller absolute values. 
Our investigation of the nonbipartite lattices (L6,L7) was, in fact, motivated by the impression that the number of interacting neighbors alone determines the entire anisotropy dependence of $\lm$.
We wanted to compare $\lm$ for the lattices (L6) and (L3), where, in the both cases, each site has four nearest neighbors, and for the lattices (L7) and (L5), each site has six nearest neighbors.

\begin{figure}
\epsfig{figure=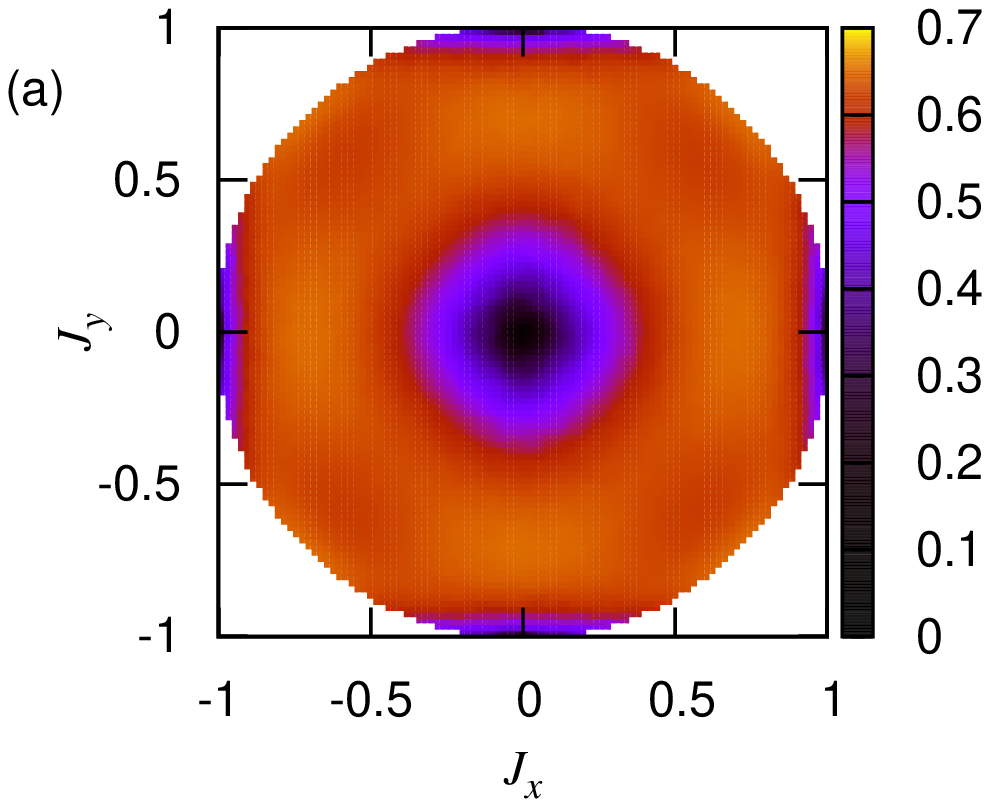,angle=270,width=6.6cm}
\epsfig{figure=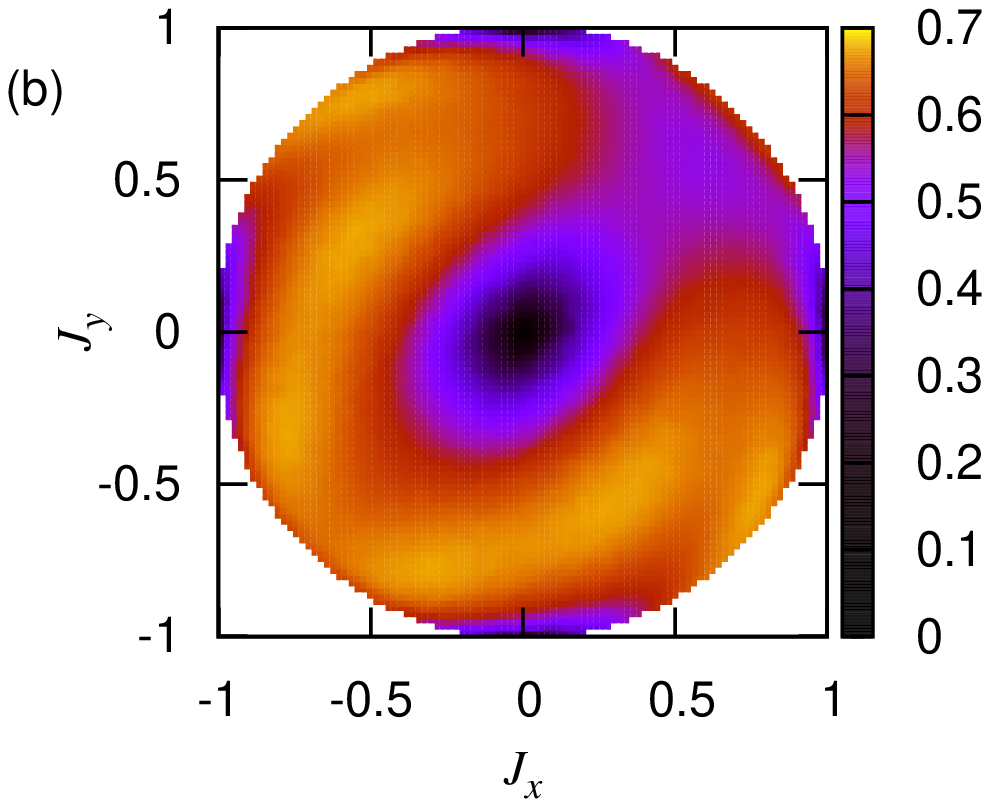,angle=270,width=6.6cm}
\caption{
(Color online) Largest Lyapunov exponent as a function of the coupling constants $J_x$ and $J_y$.
The third coupling constant is $J_z = \sqrt{1 - J_x^2 - J_y^2}$. Plots (a) and (b) represent, respectively, the numerically generated results for lattices (L5) and (L7). 
Both lattices have six nearest neighbors, but lattice (L5) is bipartite, while lattice (L7) is not. As explained in Section~\ref{Expectations}, the plots for $J_z = - \sqrt{1 - J_x^2 - J_y^2}$ can be obtained from the plots shown in this figure by reversing the signs of $J_x$ and $J_y$.
\label{fig:crosssection}
}
\end{figure}

We found, however, that, as seen in Fig.~\ref{fig:prl}, the nonbipartite lattices (L6) and (L7) exhibit a noticeable fork-like spread of $\lm$ as $\Jm$ approaches $1/\sqrt{3}$.
The upper and the lower tips of the fork correspond to the anti-Heisenberg and Heisenberg Hamiltonians, respectively.
In the anti-Heisenberg case, the value of $\lm$ is close to that of a bipartite lattice with the same number of nearest neighbors. In the Heisenberg case, the value of $\lm$ is closer to the bipartite lattice with one fewer nearest neighbor.
This spread indicates that the knowledge of $\Jm$ alone is insufficient to determine $\lm$. The two-dimensional anisotropy dependence behind this spread is shown in Fig.~\ref{fig:crosssection}, where we present it for lattices (L5) and (L7) in the form of the color density plots as a function of $J_x$ and $J_y$.

Less obvious from Fig.~\ref{fig:prl}, is the fact that the significant majority of the sampled values of $\lm$ for the lattices (L6) and (L7) agree very well with the dependence $\lm(\Jm)$ for the bipartite lattices (L3) and (L5), respectively.
The comparison between Figs.~\ref{fig:crosssection}(a) and (b) clearly shows that the difference between lattices (L5) and (L7) is only pronounced, when all three interaction constants have the same sign and roughly the same value---or, in other words, when they approach the Heisenberg limit. 
This implies that $\lm$ for the anti-Heisenberg Hamiltonian has a more typical value than for the Heisenberg Hamiltonian, and that the expected universality of $\lm(\Jm)$ for the same number of nearest neighbors still roughly holds even for nonbipartite lattices.
It thus appears that the conservation of the total spin, i.e.\ $\sum_i {\mathbf S}_i$, in the Heisenberg case leads to the reduction of the effective number of the nearest neighbors by roughly one, as far as the value of $\lm$ is concerned. 

We suspect that the situation here is similar to the origin of the frustrated low-temperature magnetism for the Heisenberg model on nonbipartite lattices.
The interaction energy for a spin pair is minimal when the two spins are antiparallel to each other, but, on a nonbipartite lattice, such an antiparallel configuration cannot simultaneously exist for all pairs of interacting spins.
Hence the ground state of a nonbipartite lattice is frustrated.
In the case of the Lyapunov instabilities, the conservation of the total spin polarization implies that
the perturbation vector ${\delta\bgamma}_1 \equiv \{ \delta {\mathbf S}_i \} $ corresponding to the largest Lyapunov exponent does not grow along the direction of the total spin polarization, i.e $\sum_i \delta {\mathbf S}_i(t) = \sum_i \delta {\mathbf S}_i(0)$. At the same time, $|\delta {\mathbf S}_i(t)|$ grows, on average, exponentially, which means that, for $t \gg 1/\lambda_1$, $|\delta {\mathbf S}_i(t)| \gg |\delta {\mathbf S}_i(0)|$. This,  in turn, implies that, in the leading order, $\sum_i \delta {\mathbf S}_i(t) \approx 0$. As a result, when a given projection $\delta {\mathbf S}_i(t)$  grows, this growth needs to be compensated by the growth of $\delta {\mathbf S}_j(t)$ for other spins in the opposite direction. However, since the interaction is local, these ``other spins'' can only be the nearest neighbors.  
Achieving the maximum growth of the perturbation, and hence the largest value of the  Lyapunov exponent presumably requires that the perturbation vector ${\delta\bgamma}_1$ maximizes the anti-alignment of  $\delta {\mathbf S}_i(t)$ for the adjacent sites.
For the bipartite lattices, this anti-alignment can, in principle, be made perfect, but for the nonbipartite lattices this is impossible.
Such an explanation is consistent with the fact that $\lm$ for lattices (L6,L7) in the Heisenberg limit is smaller than in the anti-Heisenberg limit.
It seems also to be connected to the fact that the above reduction approximately leads to the value of $\lm$ for a bipartite lattice with roughly one fewer nearest neighbor per site. 

We finally remark that the overall small spread of values of $\lm$ for bipartite lattices at a given value of $\Jm$  is related to the symmetries described in Section~\ref{Expectations}, which imply that the Lyapunov exponents are identical for eight combinations of the coupling constants characterized by the same value of $\Jm$, namely: $(J_x, J_y, J_z)$, $(-J_x, -J_y, J_z)$, $(-J_x, J_y, -J_z)$,
$(J_x, -J_y, -J_z)$, $(-J_x, -J_y, -J_z)$, $(-J_x, J_y, J_z)$, $(J_x, -J_y, J_z)$, $(J_x, J_y, -J_z)$. The values of $\lm$ cannot change much between these eight points.

\section{Summary and conclusions}

We have investigated numerically the Lyapunov spectra of systems of many classical spins for a variety of lattices and coupling constants at infinite temperature.
The possibility of varying the coupling constants, from the highly symmetric isotropic Heisenberg model, through partially symmetric couplings such as the anti-Heisenberg model, to the completely anisotropic integrable case, makes these systems particularly interesting.
We have presented: (i) calculations of the Lyapunov spectra for selected lattices of interacting classical spins; (ii) investigations of the lattice-size dependence of the Lyapunov spectra; (iii) investigations the largest Lyapunov exponents for a broader group of coupling constants, lattices and large lattice sizes; and (iv) discussion of the difference between the largest Lyapunov exponents for bipartite and nonbipartite lattices.
The computed Lyapunov spectra were found to be weakly convex.
We have observed no finite offset of the smallest positive exponents, and, to the extent that we have searched, we have not encountered any evidence of Lyapunov-Goldstone modes.
Both the largest Lyapunov exponents and the whole Lyapunov spectra were found to become independent of the lattice sizes for sufficiently large lattices. We have given an analytical argument explaining this finding.
In addition, we have found that the largest Lyapunov exponents for bipartite and nonbipartite lattices depend differently on the anisotropy of the coupling.
This is due to the special symmetry of the bipartite lattices with respect to the sign change of the two out of three coupling constants.

\section{Acknowledgements}
A.S.dW's work is financially supported by an Unga Forskare grant from the Swedish Research Council.
B.V.F. acknowledges the hospitality of the Kavli Institute for Theoretical Physics at the University of California, Santa-Barbara, where a part of this paper was written, and support by the National Science foundation under Grant No. NSF PHY11-25915. The numerical part of this work was performed at the bwGRiD computing cluster at the University of Heidelberg.

\end{document}